\def\CarlosI{Instituto de F\'\i sica Te\'orica y Computacional Carlos I, 
Facultad de Ciencias, Universidad de Granada, Campus de Fuentenueva, Granada 18002, 
Spain} 
 
\def\IAA{Instituto de Astrofisica de Andalucia, Apartado Postal 3004, Granada 
       18080, Spain} 
\def\Comision{Work partially supported by the DGICYT.} 

\def\noi{\noindent}
\def\be{\begin{equation}}
\def\ee{\end{equation}}
\def\bea{\begin{eqnarray}}
\def\eea{\end{eqnarray}}
\def\nn{\nonumber}

\documentclass[12pt]{article}

\begin{document}

\begin{center} 
{\LARGE {\bf A new attempt towards the unification of space-time and internal 
gauge symmetries$^1$ }}
\end{center} 
 
\bigskip 
\bigskip 
 
\centerline{V. Aldaya$^{2,3}$ and E. S\'anchez-Sastre$^{2,3}$}

\bigskip 
 
\footnotetext[1]{\Comision} 
\footnotetext[2]{\IAA} 
\footnotetext[3]{\CarlosI}

\bigskip 
 
\begin{center} 
{\bf Abstract} 
\end{center} 
 
\small 
\setlength{\baselineskip}{12pt} 
 
\begin{list}{}{\setlength{\leftmargin}{3pc}\setlength{\rightmargin}{3pc}} 
\item The neat 
formulation that describes the gauge interactions associated with internal symmetries 
is extended to the case 
of a simple, 
yet non-trivial, symmetry group structure which mixes gravity and electromagnetism 
by associating a gauge symmetry with a central extension of the Poincar\'e group.

\end{list} 
 
\normalsize

\noi PACS: 11.15.-q, 04.50.+h 
\setlength{\baselineskip}{14pt} 

\newpage 

\section {Introduction}

The notion of {\it gauge} 
symmetry is traced back to Weyl\cite{Weyl1} with respect to the invariance of a system under scale 
(``gauge'') transformations depending on the particular space-time point.
However, nowadays in modern physics the term ``gauge'' has nothing to do with scale transformations 
but with the whole picture that describes the fundamental interactions. 
In the standard Lagrangian formalism, promoting a given underlying rigid symmetry to 
a
``local'' one 
requires the introduction of a connection which is interpreted as a potential providing the 
corresponding gauge interaction. This is essentially the formulation of the so-called Minimal 
Coupling Principle.
Internal gauge invariance has successfully led to the electromagnetic interaction associated with 
$U(1)$, electroweak interactions associated with $(SU(2)\otimes U(1))/Z_{2}$, and finally 
to the strong interaction associated with colour $SU(3)$. As an extra bonus of gauge theory, 
the association of interactions with groups translates the problem of unification of forces
to that of finding rigid symmetry groups containing older ones as non-trivial (not as a direct 
product) subgroups. Although the final choice of a ``grand unification group'' for internal symmetry, 
of the type
$SU(5)$\cite{SU(5)} or $SO(10)$\cite{SO(10)}, still remains to be found, the actual 
problems for achieving such a result are
 of a phenomenological nature\cite{Weinberg}.

The case of the gravitational interaction understood as some sort of gauge theory is a question 
which was firstly considered by Utiyama (1956)\cite{Utiyama} and later by Kibble (1961)\cite{Kibble}. 
After these pioneer papers, much effort has been devoted to achieving a clear understanding of 
the gauge nature of the gravitational field (see among others 
$[7-31]$), although fully disconnected from other interactions. 
The unification of gravity and the other interactions would have supposedly required 
the non-trivial mixing of the space-time group and some internal symmetry, a task explicitly 
``forbidden'' long ago by the so-called ``no-go'' theorems by O'Raifeartaigh, Coleman, Mandula, 
Michel, etc. 
$[32-37]$. 
The ``no-go'' theorems 
state that there is no finite-dimensional Lie group containing the Poincar\'e 
group, acting as diffeomorphisms of the Minkowski space-time, and any 
internal $SU(n)$ group, except for the direct product.
In this paper, we shall bypass no-go theorems in a subtle way by replacing the Poincar\'e group 
with the space-time symmetry of 
the relativistic quantum particle, i.e. a central extension of the 
Poincar\'e group by $U(1)$ (see \cite{egmixing} and references there in). The proposed symmetry 
has been succesfully used in a Group Approach to Quantization (GAQ) \cite{gac}
to describe the (classical) particle-mechanics 
analog of the present problem \cite{egmixing}. GAQ was originally formulated as a group-theoretical
quantization scheme designed for obtaining the quantum dynamics of a physical system out of a given
centrally extended Lie group. However, it also describes naturally the classical limit in the 
Hamilton-Jacobi picture. 

The paper is organized as follows. Sec. II is devoted to the general structure of gauge theory including space-time symmetries. In Sec. III we present the gauging of the centrally extended Poincar\'e group giving rise to the new phenomenon of an extra coupling constant mixing non-trivially the geodesic and the Lorentz forces.



\section{Brief review of the general structure of gauge theory for internal and space-time symmetries}

\subsection{Internal symmetries} 

Let us consider a matter Lagrangian density $\mathcal{L}_{matt}(\varphi^{\alpha},\varphi_{,\mu}^{\alpha})$\footnote{The index notation throughout this paper is the 
following: we shall use the first half of the Greek alphabet $\alpha,\beta,\gamma,...(=1,...,N)$ to denote the internal components
(the representation indices) of the matter fields, the second half of the Greek alphabet $\mu,\nu,\lambda,...(=0,...,3)$ will denote 
space-time indices (the space indices running from 1 to 3 will be denoted with letters from the middle of the Latin alphabet $i,j,k,...$).
Finally we shall use the first half of the Latin alphabet in brackets $(a),(b),(c),...(=1,...,dimG)$ to denote the group indices. We emphasize that the brackets in the group indices by no means are related to symmetrization or antisymmetrization.} depending on  the matter fields $\varphi^\alpha$ and their first-order derivatives 
$\varphi_{,\mu}^\alpha \equiv \frac{\partial \varphi^\alpha}{\partial x^\mu}$.
Let us assume that the matter action

\begin{equation}
\mathcal{S}=\int\mathcal{L}_{matt}(\varphi^\alpha,\varphi^\alpha_{,\mu})
\;d^4 x
\end{equation}

\noi is invariant under a global (rigid) Lie group of internal symmetry.
The infinitesimal transformation of the matter fields (associated with each group generator with index  $(a)$)
under G is supposed to be 

\bea
\delta_{(a)}\varphi^\alpha=X^\alpha_{(a)\beta}\varphi^\alpha,
\eea
\noi where $X^\alpha_{(a)\beta}$ denotes a matrix realization of the infinitesimal action of the Lie group generators, satisfying the commutation relations

\bea
(X_{(b)}X_{(a)}-X_{(a)}X_{(b)})^{\alpha}_{\beta}=C^{c}_{ab}X^{\alpha}_{(c)\beta}, \label{ola}
\eea

\noi the $C^{c}_{ab}$ being the structure constants 
of the group.
Hence, the global invariance condition of the action reads:

 \bea
\delta^{global}_{(a)} \mathcal{L}(\varphi^\alpha,\varphi^\alpha_{,\mu})=
X^\alpha_{(a)\beta}\varphi^\beta 
\frac{\partial \mathcal{L}}{\partial \varphi^\alpha} +
X^\alpha_{(a)\beta}\varphi^\beta_{,\mu} 
\frac{\partial \mathcal{L}}{\partial \varphi^\alpha_{,\mu}}=0.
\eea








Let us consider the  (``current'', ``local'' or) gauge  
group $G(M)$, i.e. a group G with parameters depending on the space-time points. 
The corresponding Lie algebra  is the tensor product 
$\mathcal{F}(M)\otimes\mathcal{G}$
where $\mathcal{F}(M)$ is the multiplicative algebra of real analytic functions (which will be denoted in the sequel by $f^{(a)}$) on M, and 
$\mathcal{G}$ is the Lie algebra of the Lie group G. Obviously, the action is not invariant under $G(M)$:

\bea
\delta \mathcal{L}(\varphi^\alpha,\varphi^\alpha_{,\mu})=
f^{(a)} \delta^{global}_{(a)}\mathcal{L}(\varphi^\alpha,\varphi^\alpha_{,\mu})+
X^\alpha_{(a)\beta}\varphi^\beta \frac{\partial f^{(a)} (x)}{\partial x^\mu}
\frac{\partial \mathcal{L}}{\partial \varphi_{,\mu}^\alpha}=
X^\alpha_{(a)\beta}\varphi^\beta \frac{\partial f^{(a)}(x)}{\partial x^\mu}
\frac{\partial \mathcal{L}}{\partial \varphi_{,\mu}^\alpha} \neq 0.
\eea

\noi Note that $\delta^{global}_{(a)} \mathcal{L}(\varphi^\alpha,\varphi^\alpha_{,\mu})=0$ by hypothesis.
In order to restore the invariance under $G(M)$ we have to introduce new compensating fields 
(usually known 
as Yang-Mills fields) $A^{(a)}_{\mu}$ with the usual transformation law of a connection under $G(M)$: 
\begin{equation}
\delta A^{(a)}_{\mu}=f^{(b)}C^{a}_{bc}A^{(c)}_{\mu}+
{\large\frac{\partial f^{(a)}}{\partial x^{\mu}}}\,,
\end{equation}

\noi The new fields $A^{(a)}_{\mu}$  modify the 
behaviour of the original Lagrangian of matter so that we have to find,
on the one hand, the expression for the new Lagrangian $\widehat{\mathcal{L}}_{matt}$ containing the
matter fields and their interaction with the new compensating fields $A^{(a)}_{\mu}$ and, on 
the other, the free Lagrangian $\mathcal{L}_{0}$ corresponding to the new fields, which should depend 
on the new field variables and their first derivatives, i.e. $A^{(a)}_{\mu}$, $A^{(a)}_{\nu,\sigma}\equiv \frac{\partial A^{(a)}_\nu}{\partial x^\sigma}$.
It is well-known that the solution to this question is given by the Minimal Coupling Prescription, which states that {\it The new Lagrangian describing the matter fields 
as well as their interaction with the new compensating fields $A^{(a)}_{\nu}$
has the form}
\begin{equation}
\widehat{\mathcal{L}}_{matt}(\varphi^{\alpha},\varphi^{\alpha}_{,\mu},A^{(a)}_{\nu})\equiv 
\mathcal{L}_{matt}(\varphi^{\alpha},\varphi^{\alpha}_{,\mu}-A^{(a)}_{\mu}X^{\alpha}_{(a)\beta}\varphi^{\beta}).
\end{equation}

\noi In other words, the matter Lagrangian incorporating the interaction terms  is obtained from the original one by 
replacing all derivatives of the matter fields with covariant derivatives.

The introduction of the gauge (compensating) fields naturally leads to considering a new action 
accounting also for the dynamics of these new fields with Lagrange density 
$\mathcal{L}_0(A^{(a)}_\mu,A^{(b)}_{\nu,\sigma})$: 
\begin{equation}
\mathcal{S'}=\int (\widehat{\mathcal{L}}_{matt}+\mathcal{L}_{0})d^4 x.
\end{equation}

\noi Since 
$\;\int \widehat{\mathcal{L}}_{matt}\;d^4 x\;$ is 
invariant under $G(M)$, imposing the invariance of 
$\;\mathcal{S'}\;$ requires the invariance of
$\;\;\int \mathcal{L}_{0}\;d^4 x\;\;$ itself.  That is, the free Lagrangian $\mathcal{L}_{0}$, 
containing the new compensating  fields and their first derivatives, must be invariant under the 
current group $G(M)$:

\bea
\delta \mathcal{L}_{0}(A^{(c)}_{\mu},A^{(b)}_{\nu,\sigma})&=&\bigg( f^{(b)}C^{a}_{bc}A^{(c)}_{\mu}+
{\large\frac{\partial f^{(a)}}{\partial x^{\mu}}} \bigg){\large\frac{\partial \mathcal{L}_{0}}
{\partial A^{(a)}_{\mu}}}\nn\\
&+& \bigg( f^{(b)}C^{a}_{bc}A^{(c)}_{\mu,\nu}+C^{a}_{bc}A^{(c)}_{\mu}
{\large\frac{\partial f^{(b)}}{\partial x^{\nu}}}+
\frac{\partial^2f^{(a)}}{\partial x^{\nu} \partial x^{\mu}} \bigg)
{\large\frac{\partial \mathcal{L}_{0}}{\partial A^{(a)}_{\mu,\nu}}}=0\,.
\eea

\noi This requirement of gauge invariance of $\mathcal{L}_0$ implies that
{\it the necessary 
condition for $\mathcal{L}_0$ to be invariant under the current group $G(M)$ is that $\mathcal{L}_0$
depends on the fields $A^{(a)}_\mu$ and their ``derivatives'' $A^{(a)}_{\mu,\nu}$ only through the 
specific combination:}
\begin{equation}
F^{(a)}_{\mu\nu}\equiv A^{(a)}_{\mu,\nu}-A^{(a)}_{\nu,\mu}-\frac{1}{2}C^{a}_{bc}(A^{(b)}_{\mu}A^{(c)}_
{\nu}-A^{(b)}_{\nu}A^{(c)}_{\mu})\,,\label{curvatura}
\end{equation}

\noi which is traditionally called the ``curvature'' of the ``connection'' 
$A^{(a)}_\mu$ (see again the end of this subsection).

It should be remarked that the actual dependence of $\mathcal{L}_0$ on the tensor $F$ is not 
fixed and must be chosen with the help of extra criteria, for example the invariance under the rigid Poincar\'e group. In particular, to account for the 
standard Yang-Mills equations the Lagrangian must be of the form 
\begin{equation}
\mathcal{L}_{0}\sim \sum_{a=1}^{dimG} F^{(a)}_{\mu\nu}F^{(a)}_{\sigma\rho}\eta^{\sigma\mu}
\eta^{\rho\nu}\,.
\end{equation}

\noi Introducing the notation (spin connection)
\bea
\Gamma^{\alpha}_{\mu\beta}\equiv A^{(a)}_{\mu}X^{\alpha}_{(a)\beta}\,,
\eea

\noi and taking into account the commutation relations ($\ref{ola}$), the tensor $F^{(a)}_{\mu\nu}$ can be turned into a curvature tensor:
\bea
R^{\alpha}_{\mu\nu\beta}\equiv F^{(a)}_{\mu\nu}X^{\alpha}_{(a)\beta}&=&\partial_{\mu}
\Gamma^{\alpha}_{\nu\beta}-\partial_{\nu}\Gamma^{\alpha}_{\mu\beta}-\frac{1}{2}C^{a}_{bc}
(A^{(b)}_{\mu}A^{(c)}_{\nu}(X_{(a)})^{\alpha}_{\beta}-A^{(b)}_{\nu}A^{(c)}_{\mu}(X_{(a)})^{\alpha}_
{\beta})\nn\\
&=&\partial_{\mu}\Gamma^{\alpha}_{\nu\beta}-\partial_{\nu}\Gamma^{\alpha}_{\mu\beta}-
(\Gamma^{\alpha}_{\mu\gamma}\Gamma^{\gamma}_{\nu\beta}-\Gamma^{\alpha}_{\nu\beta}\Gamma^{\gamma}_
{\mu\beta})\,.
\eea

\noi The content of this subsection summarizes briefly the general scheme of
the well-known formulation of gauge theory associated with internal symmetry groups. Subtle questions such as the Higgs-Kibble mechanism via  spontaneous symmetry breaking are not considered in the present paper. However, in a forthcoming work \cite{jetgroups}
we shall propose an alternative mass-generating mechanism for the gauge vector bosons which is based essentially on the introduction of the group parameters
 in the theory as dynamical fields.  

\subsection{Space-time symmetries}

In this subsection we  generalize the previous one to the case in which the rigid group 
also acts on the space-time. The infinitesimal transformation of the space-time coordinates and the matter fields is taken to be of the form

\bea
\delta_{(a)}x^\mu&=&X^\mu_{(a)}\\
\delta_{(a)}\varphi^\alpha&=&X^\alpha_{(a)\beta}\varphi^\beta,
\eea

\noi where  $X^\mu_{(a)}$ is in general a function of the position. As in the internal symmetry case the starting point of the theory is the hypothesis of global invariance of the matter action, i.e.

\bea
X^\mu_{(a)}\frac{\partial \mathcal{L}_{matt}}{\partial x^\mu}+
X^{\alpha}_{(a)\beta}\varphi^\beta \frac{\partial \mathcal{L}_{matt}}{\partial \varphi^\alpha}+
(X^{\alpha}_{(a)\beta}\varphi^\beta_{,\mu}-\varphi^\alpha_{,\nu}\frac{\partial X^\nu_{(a)}}{\partial x^\mu})
\frac{\partial \mathcal{L}_{matt}}{\partial \varphi^\alpha_{,\mu}}
+
\mathcal{L}_{matt}\partial_\mu X^\mu_{(a)}=0.\label{variaciondens}
\eea

\noi It is remarkable the appearance of the divergence of the action of the group on the space-time 
coordinates $\partial_\mu X^\mu_{(a)}$, a term which was absent for the internal symmetry case. This is a 
consequence of the variation of the integration volume: $\delta_{(a)}d^4x=\partial_\mu X^\mu_{(a)}d^4x$.

Let us construct an invariant action under the local (gauge) space-time group generated by:

\bea
f^{(a)}(x)\delta_{(a)}x^\mu&=&f^{(a)}(x)X^\mu_{(a)}\\
f^{(a)}(x)\delta_{(a)}\varphi^\alpha&=&f^{(a)}(x)X^\alpha_{(a)\beta}\varphi^\beta.
\eea

\noi It is worth realizing that the dependence of the space-time components of the generators on $x^\mu$ 
through $X^\mu_{(a)}$ is not ``gauge''. The gauge dependence on $x^\mu$ arises from the fact that these 
generators are multiplied by arbitrary functions $f^{(a)}(x)$.

The construction of a gauge invariant Lagrangian density requires the introduction of new fields. 
Apart from compensating fields $\mathcal{A}^{(a)}_{\nu}$ analogous to those of internal symmetries, 
there will be additional compensating fields $k^\nu_\mu$ (tetrad fields) related to the group 
action on the space-time. The corresponding transformation laws of the compensating fields under $G(M)$
 read:
\bea
\delta \mathcal{A}^{(a)}_{\mu}&=&f^{(b)}C^{a}_{bc}
\mathcal{A}^{(c)}_{\mu}+k^{\nu}_{\mu}{\large\frac{\partial f^{(a)}}{\partial x^{\nu}}}-f^{(b)}
\mathcal{A}^{(a)}_{\sigma}{\large\frac{\partial X^{\sigma}_{(b)}}{\partial x^{\mu}}}\\
\delta k^{\nu}_{\mu}&=&X^{\nu}_{(a)}k^{\sigma}_{\mu}{\large\frac{\partial 
f^{(a)}}{\partial x^{\sigma}}}+f^{(a)}\bigg( k^{\sigma}_{\mu}{\large\frac{\partial X^{\nu}_{(a)}}
{\partial x^{\sigma}}}-k^{\nu}_{\sigma}{\large\frac{\partial X^{\sigma}_{(a)}}{\partial x^{\mu}}} \bigg).\label{deltak}
\eea

\noi Inverse fields of $k^{\nu}_{\mu}$ will be denoted by $q^{\mu}_{\sigma}$, so that

\bea
k^{\nu}_{\mu}q^{\mu}_{\sigma}&=&\delta^{\nu}_{\sigma}\\
k^{\nu}_{\mu}q^{\sigma}_{\nu}&=&\delta^{\sigma}_{\mu}.
\eea

\noi Following similar steps to those in the internal case we can establish a generalized Minimal 
Coupling Prescription by saying that {\it the new Lagrangian describing the matter fields 
as well as their interaction with the compensating fields $\mathcal{A}^{(a)}_{\nu}$, $k^{\nu}_{\mu}$ 
has the form}
\begin{equation}
\widehat{\mathcal{L}}_{matt}(\varphi^{\alpha},\varphi^{\alpha}_{,\mu},\mathcal{A}^{(a)}_{\nu},
k^{\nu}_{\mu})\equiv \mathcal{L}_{matt}(\varphi^{\alpha},k^{\nu}_{\mu}\varphi^{\alpha}_{,\nu}-
\mathcal{A}^{(a)}_{\mu}X^{\alpha}_{(a)\beta}\varphi^{\beta}),
\end{equation}

\noi although the expression $k^{\nu}_{\mu}\varphi^{\alpha}_{,\nu}-
\mathcal{A}^{(a)}_{\mu}X^{\alpha}_{(a)\beta}\varphi^{\beta}$ can no longer be considered as a 
covariant derivative. Let us prove the gauge invariance of the action associated with this Lagrangian, 
i.e. let us see that

\bea
\delta \hat{S}_{matt}=0
\eea

\noi where

\bea
\hat{S}_{matt}&=&\int \hat{L}_{matt}\,d^4 x\equiv\int \Lambda\widehat{\mathcal{L}}_{matt}(\varphi^{\alpha},\varphi^{\alpha}_{,\mu},\mathcal{A}^{(a)}_{\nu},
k^{\nu}_{\mu})\,d^4 x\nn\\
&=&\int \Lambda \mathcal{L}_{matt}(\varphi^{\alpha},k^{\nu}_{\mu}\varphi^{\alpha}_{,\nu}-
\mathcal{A}^{(a)}_{\mu}X^{\alpha}_{(a)\beta}\varphi^{\beta})\,d^4 x,
\eea

\noi the factor $\Lambda$ being a function of the tetrad fields to be determined by demanding the gauge invariance of
$\hat{S}_{matt}$. 
The infinitesimal variation of $\widehat{\mathcal{L}}_{matt}(\varphi^{\alpha},\varphi^{\alpha}_{,\mu},\mathcal{A}^{(a)}_{\nu},k^{\nu}_{\mu})$ under $G(M)$ reads:

\bea
\delta \widehat{\mathcal{L}}_{matt}&=&f^{(a)}X^\mu_{(a)}\frac{\partial \widehat{\mathcal{L}}_{matt}}{\partial x^\mu}+
f^{(a)}X^{\alpha}_{(a)\beta}\varphi^{\beta}\frac{\partial \widehat{\mathcal{L}}_{matt}}{\partial \varphi^\alpha} \nn\\
&+&(\partial_\mu f^{(a)}X^{\alpha}_{(a)\beta}\varphi^\beta-\varphi^\alpha_{,\nu}(\partial_\mu f^{(a)}X^\nu_{(a)}+
f^{(a)}\partial_\mu X^\nu_{(a)})+f^{(a)}X^\alpha_{(a)\beta}\varphi^\beta_{,\mu})
\frac{\partial \widehat{\mathcal{L}}_{matt}}{\partial \varphi^\alpha_{,\mu}}\nn\\
&+&(f^{(b)}C^a_{bc}\mathcal{A}^{(c)}_{\mu}+k^\nu_\mu \partial_\nu f^{(a)}-f^{(b)}\mathcal{A}^{(a)}_{\sigma}\partial_{\mu}X^{\sigma}_{(b)})
\frac{\partial \widehat{\mathcal{L}}_{matt}}{\partial \mathcal{A}^{(a)}_{\mu}}\nn\\
&+&(X^\nu_{(a)}k^\sigma_\mu \partial_\sigma f^{(a)}+f^{(a)}(k^\sigma_\mu \partial_\sigma X^\nu_{(a)}-
k^\nu_\sigma \partial_\mu X^\sigma_{(a)}))\frac{\partial \widehat{\mathcal{L}}_{matt}}{\partial k^\nu_\mu}.\label{variacionmatt} 
\eea

\noi Let us consider the following change of variables:

\bea
\phi^{\alpha}&=&\varphi^{\alpha}\nn\\
\phi^{\alpha}_{,\mu}&=&k^{\nu}_{\mu}\varphi^{\alpha}_{,\nu}-\mathcal{A}^{(a)}_{\mu}X^{\alpha}_{(a)\beta}\varphi^{\beta}\nn\\
{\cal B}^{(a)}_{\mu}&=&\mathcal{A}^{(a)}_{\mu}\label{transformacion}\\
K^{\mu}_{\nu}&=&k^{\mu}_{\nu}\nn\\
Q^{\mu}_{\nu}&=&q^{\mu}_{\nu},\nn
\eea

\noi and the corresponding change in the partial derivatives:
\bea
{\large\frac{\!\partial }{\partial \varphi^{\alpha}}}&=&{\large\frac{\!\partial }
{\partial \phi^{\alpha}}}-{\cal B}^{(a)}_{\mu}X^{\beta}_{(a)\alpha}{\large\frac{\!\partial }{\partial 
\phi^{\beta}_{,\mu}}}\nn\\
{\large\frac{\!\partial }{\partial \varphi^{\alpha}_{,\mu}}}&=&K^{\mu}_{\nu}{\large\frac{\!\partial }
{\partial \phi^{\alpha}_{,\nu}}}\nn\\
{\large\frac{\!\partial }{\partial \mathcal{A}^{(a)}_{\mu}}}&=&{\large\frac{\!\partial }
{\partial {\cal B}^{(a)}_{\mu}}}-X^{\alpha}_{(a)\beta}\phi^{\beta}{\large\frac{\!\partial }
{\partial \phi^{\alpha}_{,\mu}}}\\
{\large\frac{\!\partial }{\partial k^{\nu}_{\mu}}}&=&{\large\frac{\!\partial }{\partial K^{\nu}_{\mu}}}+
Q^{\sigma}_{\nu}(\phi^{\alpha}_{,\sigma}+{\cal B}^{(a)}_{\sigma}X^{\alpha}_{(a)\beta}\phi^{\beta}){\large\frac{\!\partial }
{\partial \phi^{\alpha}_{,\mu}}}.\nn
\eea

\noi With the help of this change of variables the infinitesimal variation (\ref{variacionmatt})
under the local space-time symmetry group can be written as $f^{(a)}$ times the global variation 
of the original matter Lagrangian density of the theory depending on the field variables 
$\phi^\alpha$ and $\phi^\alpha_{,\mu}$, i.e.

\bea
\delta \widehat{\mathcal{L}}_{matt}(\varphi^\alpha,\varphi^\alpha_{,\mu},\mathcal{A}^{(a)}_{\nu},
k^\nu_\mu)=f^{(a)}\delta_{(a)}^{global}\mathcal{L}_{matt}(\phi^\alpha, \phi^\alpha_{,\mu})
\eea

where 

\bea 
\delta_{(a)}^{global}\mathcal{L}_{matt}(\phi^\alpha, \phi^\alpha_{,\mu})\equiv X^\nu_{(a)}
\frac{\partial \mathcal{L}_{matt}}{\partial x^\nu}+
X^{\gamma}_{(a)\beta}\phi^\beta \frac{\partial \mathcal{L}_{matt}}{\partial \phi^\gamma}+
(X^{\gamma}_{(a)\beta}\phi^\beta_{,\nu}-\phi^\gamma_{,\sigma}\frac{\partial X^\sigma_{(a)}}{\partial x^\nu})
\frac{\partial \mathcal{L}_{matt}}{\partial \phi^\alpha_{,\nu}}.
\eea

\noi Using the hypothesis of invariance of the matter action under the global group (see (\ref{variaciondens})) it follows that

\bea
\delta \widehat{\mathcal{L}}_{matt}(\varphi^\alpha,\varphi^\alpha_{,\mu},\mathcal{A}^{(a)}_{\nu},
k^\nu_\mu)&=&-f^{(a)}\mathcal{L}_{matt}(\phi^\alpha, \phi^\alpha_{,\mu})\partial_\mu X^\mu_{(a)}\nn\\
&=&-f^{(a)}\mathcal{L}_{matt}(\varphi^\alpha, k^\nu_\mu\varphi^\alpha_{,\nu}-\mathcal{A}^{(a)}_\mu X^\alpha_{(a)\beta}\varphi^\beta)\partial_\mu X^\mu_{(a)}. \label{var2}
\eea

\noi Let us determine the simplest form of the factor $\Lambda$ that leads to a gauge invariant 
Lagrangian density $\widehat{L}_{matt}\equiv \Lambda \widehat{\mathcal{L}}_{matt}$. 
$\widehat{L}_{matt}$ must satisfy the condition:

\bea
\delta \widehat{L}_{matt}+\widehat{L}_{matt}\partial_\mu (f^{(a)}X^\mu_{(a)})=0.\label{var3}
\eea

\noi More explicitly,

\bea
\delta \Lambda \widehat{\mathcal{L}}_{matt}+\Lambda \delta \widehat{\mathcal{L}}_{matt}
+\Lambda \widehat{\mathcal{L}}_{matt}\partial_\mu f^{(a)}X^\mu_{(a)}+\Lambda \widehat{\mathcal{L}}_{matt}f^{(a)}\partial_\mu X^\mu_{(a)}=0\,.
\eea

\noi Assuming that 
\bea
\widehat{\mathcal{L}}_{matt}(\varphi^\alpha,\varphi^\alpha_{,\mu},\mathcal{A}^{(a)}_{\nu},
k^\nu_\mu)=\mathcal{L}_{matt}(\varphi^\alpha,k^{\nu}_{\mu}\varphi^{\alpha}_{,\nu}-
\mathcal{A}^{(a)}_{\mu}X^{\alpha}_{(a)\beta}\varphi^{\beta})
\eea

\noi  and using (\ref{var2}), the gauge invariance condition of $\widehat{L}_{matt}$ (\ref{var3}) 
provides the equation that $\Lambda$ must satisfy, that is:

\bea
\delta \Lambda +\Lambda \partial_\mu f^{(a)}X^\mu_{(a)}=0.
\eea

\noi For simplicity we shall assume that $\Lambda$ only depends on the tetrad fields, so that

\bea
\delta \Lambda=\frac{\partial \Lambda}{\partial k^\nu_\mu}\delta k^\nu_\mu\,,
\eea

\noi and taking into account (\ref{deltak}) the final equation that determines the form of 
$\Lambda$ reads:

\bea
(X^{\nu}_{(a)}k^{\sigma}_{\mu}
\partial_\sigma f^{(a)}
+f^{(a)}( k^{\sigma}_{\mu}\partial_\sigma X^{\nu}_{(a)}
-k^{\nu}_{\sigma}\partial_\mu X^{\sigma}_{(a)}))
\frac{\partial \Lambda}{\partial k^\nu_\mu}+\Lambda \partial_\mu f^{(a)}X^\mu_{(a)}=0.
\eea

\noi Since the functions $f^{(a)}$ are arbitrary and independent, the coefficients of $f^{(a)}$ and their first-order derivatives must be zero, so that we obtain the following system of partial differential equations:

\bea
&a)&\;\;f^{(a)}:( k^{\sigma}_{\mu}\partial_\sigma X^{\nu}_{(a)}
-k^{\nu}_{\sigma}\partial_\mu X^{\sigma}_{(a)})
\frac{\partial \Lambda}{\partial k^\nu_\mu}=0\\
&b)&\;\;\partial_\sigma f^{(a)}:X^{\nu}_{(a)}k^{\sigma}_{\mu}\frac{\partial \Lambda}{\partial k^\nu_\mu}+\Lambda X^\sigma_{(a)}=0,
\eea

\noi and the general solution for this system is

\bea
\Lambda=det(q^\nu_\mu).
\eea

\noi Note that when $k^\nu_\mu\rightarrow \delta^\nu_\mu$ (internal symmetry case) then $\Lambda\rightarrow 1$.
As a corollary, we can assert that {\it the new  action invariant under the local space-time symmetry  
describing the matter fields as well as their interaction with the compensating (gauge) 
fields $\mathcal{A}^{(a)}_{\nu}$, $k^{\nu}_{\mu}$ reads}
\begin{equation}
\hat{S}_{matt}=\int \hat{L}_{matt}\,d^4 x\equiv\int \Lambda\widehat{\mathcal{L}}_{matt}\,d^4 x,
\end{equation}

\noi where $\Lambda\equiv det(q^{\nu}_{\mu})$.

If we introduce new ``tetrad-like'' compensating fields $h^{(a)\nu}_{\mu\sigma}$ associated with each generator by means of the decomposition of the tetrad field

\bea
k^\nu_\mu=\delta^\nu_\mu-h^{(a)\nu}_{\mu\sigma}X^\sigma_{(a)}
\eea

\noi we can write the interaction term in the way 

\bea
\varphi^{\alpha}_{,\mu}-\mathcal{A}^{(a)}_{\mu}X^{\alpha}_{(a)\beta}\varphi^{\beta}-h^{(a)\nu}_{\mu\sigma}X^\sigma_{(a)}\varphi^\alpha_{,\nu}
\eea

\noi that generalizes more directly the case of internal symmetry (a similar expression was already 
suggested in a footnote in \cite{Kibble}). 
>From this expression we can observe that while the gauge potentials
associated with the internal action of the group couple to the matter fields, the fields $h^{(a)\nu}_{\mu\sigma}$ couple to the derivatives of the matter
fields.
Note also that the two indices of $k^\nu_\mu$ transform according to different transformation rules, i.e. while the index $\nu$ transforms as a tensor, the index $\mu$ inherits the non-tensorial character of $h^{(a)\nu}_{\mu\sigma}$. We shall not make an explicit distinction in the notation for the tetrad indices. 
No confusion should arise since tetrads ($k$) and their inverse ($q$) are denoted differently.

%


As far as the Lagrangian $\mathcal{L}_0$ for the free compensating fields
is concerned we can establish the following theorem:
{\it The necessary condition for $\mathcal{L}_0$ to be invariant under the current group $G(M)$ is that 
$\mathcal{L}_0$ depends on the fields $\mathcal{A}^{(a)}_\mu,\,k^\nu_\mu$ and their ``derivatives'' 
$\mathcal{A}^{(a)}_{\mu,\nu}\,,\;k^\nu_{\mu,\sigma}$ only through the 
specific combination (generalized ``curvature"):}
\[
{\mathcal F}^{(a)}_{\mu\nu}\equiv {\cal A}^{(a)}_{\mu,\sigma}k^{\sigma}_{\nu}-
{\mathcal A}^{(a)}_{\nu,\sigma}k^{\sigma}_{\mu}-\frac{1}{2}C^{a}_{bc}({\cal A}^{(b)}_{\mu}{\cal A}^{(c)}_{\nu}-
{\mathcal A}^{(b)}_{\nu}{\cal A}^{(c)}_{\mu})-{\cal A}^{(a)}_{\sigma}T^{\sigma}_{\mu\nu}\,,
\]

\noi with $T^{\sigma}_{\mu\nu}\equiv q^{\sigma}_{\rho}(k^{\rho}_{\mu,\tau}k^{\tau}_{\nu}- k^{\rho}_{\nu,\tau}k^{\tau}_{\mu})$.

The gauge-invariant action for the compensating fields has the form

\bea
\widehat{S}_0=\int \widehat{L}_0\,d^4 x\equiv\int 
\Lambda\widehat{\cal L}_0\,d^4 x.
\eea

\noi Now that a generalized gauge theory including  space-time symmetries is available, different 
gauge gravitational theories can be constructed using several space-time symmetry groups: 
space-time translations, Lorentz group, Poincar\'e group, Weyl group, etc. and the resulting 
theories can be reduced to the Einstein's theory in some particular cases.
As an example, and since we shall be concerned with
the Poincar\'e group in the next section, the rest of this subsection will be devoted to the 
gravitational theory associated with the gauge theory of the
 Poincar\'e group (see \cite{Kibble} 
among others).

The notation for the Poincar\'e group (semidirect product of the translations group and Lorentz group) index is  $(a)=\{(\mu)\;\hbox{translations},\,(\nu\sigma)\;\hbox{Lorentz}\}$ and a particular realization for the generators of the rigid
Poincar\'e algebra reads:\\

\noi Translations: 
 
\bea
\delta_{(\mu)} x^\nu&=&\delta^{\nu}_{\mu}\\
\delta_{(\mu)} \varphi^\alpha&=&0
\eea

\noi Lorentz: 

\bea
\delta_{(\mu\nu)}x^\sigma&=&\delta^{\sigma}_{(\mu\nu),\rho}x^{\rho}\equiv (\delta^{\sigma}_{\mu}\eta_{\nu\rho}-\delta^{\sigma}_{\nu}\eta_{\mu\rho})x^{\rho}\\
\delta_{(\mu\nu)}\varphi^\alpha&=&S_{(\mu\nu)\beta}^{\alpha}\varphi^{\beta}
\eea

\noi and the form of  $S_{(\mu\nu)\beta}^{\alpha}$ is determined by the commutation relations of the Poincar\'e group and antisymmetry in the Lorentz indices $S_{(\mu\nu)\beta}^{\alpha}=-S_{(\nu\mu)\beta}^{\alpha}$.


In the present case, the Lagrangian for the free compensating fields 
$\mathcal{A}^{(\nu)}_\mu$, $\mathcal{A}^{(\nu\sigma)}_\mu$, $k^\nu_\mu$
is an arbitrary function of the 
translational and Lorentz generalized curvatures, according to the previous general theory of 
gauged space-time algebras,
\begin{equation}
\mathcal{L}_{0}=\mathcal{L}_{0}(\mathcal{F}^{(\mu)}_{\nu\sigma},\,\mathcal{F}^{(\nu\sigma)}_{\rho\theta}).
\end{equation}

\noi As a particular case we can choose

\begin{equation}
\mathcal{L}_{0}=\mathcal{L}_{0}(\mathcal{F}^{(\nu\sigma)}_{\rho\theta}).
\end{equation}

\noi and by means of the decomposition of the tetrad fields in terms of the translational gauge fields

\bea
k^\mu_\nu=\delta^\mu_\nu+\mathcal{A}^{(\mu)}_\nu
\eea

\noi one can obtain (by combining the equations of motion associated with
$\mathcal{A}^{(\mu)}_\nu$ and $k^\nu_\mu$) 
the following generalized Einstein's equation: 
\begin{equation}
\mathcal{F}^{(\sigma\rho)}_{\mu\nu}\frac{\partial L_{0}}{\partial \mathcal{F}^{(\sigma\rho)}_{\epsilon\nu}}-
\frac{1}{2}\delta^{\epsilon}_{\mu}L_{0}=-\frac{1}{2}k^{\epsilon}_{\xi}t^{\xi}_{\mu}\,,
\end{equation}

\noi where

\bea
L_0&\equiv& \Lambda \mathcal{L}_0(\mathcal{F}^{(\nu\sigma)}_{\rho\theta}),\\
t^\mu_\nu&\equiv& q^{\mu}_{\sigma}\left(-\delta^{\sigma}_{\nu}\widehat{L}_{matt}+\frac{\partial \widehat{L}_{matt}}
{\partial \varphi^{\alpha}_{,\sigma}}\phi^{\alpha}_{\rho}q^{\rho}_{\nu}\right),\\
\phi^{\alpha}_{\rho}&\equiv& k^{\nu}_{\rho}\varphi^{\alpha}_{,\nu}-\mathcal{A}^{(a)}_{\rho}X^{\alpha}_
{(a)\beta}\varphi^{\beta},\\
\widehat{L}_{matt}&=&\Lambda \mathcal{L}_{matt}(\varphi^\alpha,\phi^\alpha_\mu).
\eea

Let us consider two cases:

\bigskip

\noi {\bf A) Equations in vacuum:} The action reduces to that for the free compensating fields,

\begin{equation}
\mathcal{S}_{0}=\int(\Lambda\mathcal{L}_{0})d^4 x\,,
\end{equation}

\noi and, with the choice
$\mathcal{L}_{0}=\mathcal{F}^{(\mu\nu)}_{\sigma\rho}\eta^{\sigma}_{\mu}\eta^{\rho}_{\nu}$,
the equation of motion $\frac{\delta L_{tot}}{\delta \mathcal{A}^{(a)}_{\mu}}=0$ yields
\be
\mathcal{A}_{(\sigma\rho)\mu}=\frac{1}{2}T_{\mu\sigma\rho}+\frac{1}{2}(T_{\sigma\rho\mu}-T_{\rho\sigma\mu})\,,\label{vacuumA}
\ee

\noi where 
$
\mathcal{A}_{(\sigma\rho)\mu}\equiv \mathcal{A}^{(\theta\epsilon)}_{\mu}\eta_{\theta\sigma}\eta_{\epsilon\rho}$
 and $T_{\mu\sigma\rho}\equiv T^{\nu}_{\sigma\rho}\eta_{\nu\mu}\,.$
Note that $\mathcal{A}^{(\sigma\rho)}_{\mu}=-\mathcal{A}^{(\rho\sigma)}_{\mu}$ and $T^{\nu}_{\sigma\rho}=
-T^{\nu}_{\rho\sigma}$.
Substituting (\ref{vacuumA})
into $L_{0}$, one easily finds that the theory reduces to Einstein's vacuum theory.

\bigskip 
\noi {\bf B) Equations with matter:} In this case the total action must include a matter piece 
which should be made explicit. Then, only general comments can be pointed out. For instance, 
the expression (\ref{vacuumA}) now reads
\begin{equation}
\mathcal{A}_{(\sigma\rho)\mu}=\frac{1}{2}T_{\mu\sigma\rho}+\frac{1}{2}(T_{\sigma\rho\mu}-T_{\rho\sigma\mu})+M_{(\sigma\rho)\mu}\,,
\end{equation}
 
\noi where the extra term 
$M_{(\sigma\rho)\mu}$ is zero for spinless matter but not for fermionic matter. Then, 
for a Dirac spinor $\psi$, $M_{(\sigma\rho)\mu}$  is proportional to 
$\bar{\psi}\gamma_{\mu}\Sigma_{\sigma\rho}\psi$ and this term is known as the {\it contortion} 
created by spinors \cite{Cho2}. See also \cite{Kibble}.

This situation generalizes Einstein's theory with a Lagrangian density, 
\begin{equation}
\Lambda \mathcal{F}^{(\mu\nu)}_{\mu\nu}=\Lambda R^{(\Gamma_{Levi-Civita})}+
\Upsilon (M_{(\sigma\rho)\mu})\,,
\end{equation}
 
\noi where the form of the function $\Upsilon$ again depends on the specific nature of 
fermionic matter.

Had $\mathcal{L}_{0}$ depended also on $\mathcal{F}^{(\mu)}_{\nu\sigma}$ we would have obtained a 
theory even more general than 
Einstein's, known as Einstein-Cartan theory, in which $\mathcal{F}^{(\mu\nu)}_{\sigma\rho}$ 
is interpreted as a curvature and $\mathcal{F}^{(\mu)}_{\nu\sigma}$ as a torsion.\\

We would like to remark that the formulation of the gauge theory associated with space-time symmetry 
groups, which has been presented in this subsection, not only can be applied to groups higher than the 
Poincar\'e group (in this sense this theory  would be more general than that of \cite{Kibble}) 
as for instance the Weyl group, but also, this framework results specially suitable 
for the unification of interactions. The crucial point is the incorporation of non-trivial gauge 
translational potentials even though the corresponding generators do not act on the internal 
components of the matter fields.

\section{Towards a mixing of gravity and electromagnetism}

The present section is devoted to a simple, yet non-trivial framework to account for the mixing of 
gravitation and the rest of fundamental interactions. In our approach we make use of two important 
physical notions: the well-known gauge invariance principle and the concept of central extension of 
a group (in particular, the central extension of Poincar\'e group, $\mathcal{P}$, by $U(1)$, denoted 
in the following as $\tilde{\mathcal{P}}$). On the one hand, the gauge invariance is the key for the understanding of the formulation of the interactions and
is a requirement 
that helps to achieve renormalizability. 
 Moreover, the interest 
in the description of gravity as a gauge theory is precisely the possibility of its unification with 
the rest of interactions. On the other hand, the motivation for considering a centrally extended 
group is based on the relevance of this notion in some areas of physics, specially in quantum theory
(also in classical mechanics in the Hamilton-Jacobi approach). In fact, traditional space-time 
groups as Galilei or Poincar\'e groups leave only semi-invariant the Lagrangians of the corresponding
free particles, and a central extension is required to achieve strict invariance. It is also well 
known that the Schr\"odinger equation for the free particle is not invariant under the Galilei group 
$G$ although it is under the centrally 
extended Galilei group $\tilde{G}_{(m)}$\footnote{The particular case of central extensions of Lie groups by $U(1)$ (whose classification was carried 
out long ago by Bargmann \cite{bargmann}) is very important from the physical point of view. In fact, it is known 
that the question of the classification of all the possible
projective unitary representations of a group (which are the relevant representations in quantum 
mechanics) is equivalent to the problem of the classification of the central extensions of a group 
by $U(1)$.}. Analogously, we can 
consider the space-time symmetry of the quantum relativistic particle, 
which is characterized by the commutator of boosts and translations modified with the central 
generator $\Xi$ associated with $U(1)$, i.e.

\begin{equation}
[K^{i},P_{j}]=\delta^{i}_{j}\left(\frac{1}{c}P_{0}+\lambda^{0}\Xi\right)\,. \label{non-covariant}
\end{equation}

\noi with $\lambda^{0}\equiv m$. 
In this case, a particular four-vector $\lambda$ of the orbit $\lambda^{2}=m^{2}$
in the momentum space
has been chosen. In the non-relativistic limit this commutator
yields the basic commutators of the centrally extended Galilei group, $\widetilde{G}_{(m)}$.

In the present paper we shall approach the mixing between electromagnetism and gravity 
by studying the gauge symmetry of the central extension of Poincar\'e group by $U(1)$, denoted by 
$\tilde{\mathcal{P}}$.  The group index $(a)$ now runs over $\{(\mu)$ translation, $(\nu\sigma)$ 
Lorentz, $(\Phi)$ $U(1)\}$.
The commutator of Lorentz and translations generators is modified according to

\bea
[\widetilde{M}_{\mu\nu},\widetilde{P}_{\rho}]=\eta_{\nu\rho}\widetilde{P}_{\mu}-\eta_{\mu\rho}\widetilde{P}_{\nu}-(\lambda_{\mu}\eta_{\nu\rho}-\lambda_{\nu}\eta_{\mu\rho})\Xi\equiv C^{\sigma}_{\mu\nu,\rho}\widetilde{P}_{\sigma}+C^{\Phi}_{\mu\nu,\rho}\Xi,
\eea

\noi with

\bea
C^{\Phi}_{\mu\nu,\rho}\equiv \lambda_{\nu}\eta_{\mu\rho}-\lambda_{\mu}\eta_{\nu\rho},
\eea

\noi where $\Xi$ is the generator of $U(1)$ and $\lambda_{\mu}$ is a vector in the Poincar\'e coalgebra belonging to a given coadjoint orbit, and will be related later to the coupling constant of the mixing.

\noi From the strict mathematical point of view, the group $\widetilde{\mathcal{P}}$ is a trivial central 
 extension of the Poincar\'e group by $U(1)$.
In fact, by making the replacement

\bea
P_{\mu} \rightarrow \widetilde{P}_{\mu}=P_{\mu}+\lambda_{\mu}\Xi
\eea

\noi it becomes clear that $\widetilde{\mathcal{P}}$ is equivalent to 
$\mathcal{P} \otimes U(1)$. Therefore the associated co-cycle is trivial, i.e. co-boundary. It is known, however, that 
trivial co-cycles can be divided into two different types depending on the structure of their 
generating functions \cite{Saletan}. The first type comprises the co-boundaries which are really physically
trivial as they lead to zero curvature. The second type (and the truly relevant from the physical point of view)
corresponds to those co-boundaries leading to a group connection with non-trivial curvature and are called 
pseudo-co-cycles. The central extensions that they provide are referred to as pseudo-extensions. The most remarkable fact is
that non-trivial symplectic structures and dynamics can be derived out of them 
\cite{aldazca,aldana}. An example of pseudo-extension is the case 
of the central extension of the Poincar\'e group by $U(1)$. As we shall see in the present section, this group is associated with a gauge symmetry which in particular
generates
a $U(1)$-field 
strength (containing terms of pure gravitational origin) that is not trivial as a consequence of the associated co-boundary being a pseudo-co-cycle.\footnote{The characterization of the classes of pseudo-extensions associated with non-equivalent symplectic structures leads to the notion of
pseudo-cohomology. As a report on pseudo-extensions, and the role that they play in representation theory, we refer the reader to 
Ref. \cite{Pseudo} and references there in. Here we would like to mention briefly some indications of the need of
pseudo-cohomology. It is known that pseudo-co-cycles play a fundamental role in representation theory of  
semi-simple groups (including infinite-dimensional ones like $Diff(S^1)$ and other 
diffeomorphism groups) and also in the explicit construction of the local exponent associated with Lie algebra co-cycles of the
 corresponding Kac-Moody groups \cite{aldana,Mickelsson}. In any case, the framework where the need and relevance of pseudo-cohomology is more patent is the 
so-called Group Approach to Quantization (GAQ) (mentioned in the Introduction).}


Let us consider the gauge theory of $\widetilde{\mathcal{P}}$. 
Proceeding according to the general theory developed in the subsection 2.2, the Lagrangian for the free compensating fields should be a general function of the generalized curvatures: 

\begin{equation}
\mathcal{L}_{0}=\mathcal{L}_{0}(\mathcal{F}^{(\sigma)}_{\mu\nu},\,\mathcal{F}^{(\sigma\rho)}_{\mu\nu},
\,\mathcal{F}^{(\Phi)}_{\mu\nu})\,.
\end{equation}

\noi Let us define the fields $A^{(a)}_{\mu}\equiv q^{\nu}_{\mu}\mathcal{A}^{(a)}_{\nu}$ and write the curvatures in the following way:

\bea
\mathcal{F}^{(a)}_{\mu\nu}\equiv k^{\sigma}_{\mu}k^{\rho}_{\nu}F^{(a)}_{\sigma\rho}\,,
\eea

\noi where
\bea
F^{(a)}_{\sigma\rho}\equiv A^{(a)}_{\sigma,\rho}-A^{(a)}_{\rho,\sigma}+\frac{1}{2}\widetilde{C^{a}_{bc}}(A^{(b)}_{\sigma}A^{(c)}_{\rho}-A^{(b)}_{\rho}A^{(c)}_{\sigma})\,.
\eea


\noi Here, $(a)$ runs over the entire group $\widetilde{\mathcal{P}}$ and $\widetilde{C^{a}_{bc}}$
denotes its structure constants.  The presence of a coupling constant of the mixing,  $\kappa$, through $C^{\Phi}_{\mu,\sigma\rho}$
in the generalized curvature $\mathcal{F}^{(\Phi)}_{\mu\nu}$, due to the central pseudo-extension, is to be 
remarked. In fact, and without loss of generality we can select a preferred direction 
for $\lambda_{\mu}$,

\bea
\lambda_{\mu}=-\kappa \delta^{0}_{\mu},
\eea

\noi so that we arrive at
\be
C^{\Phi}_{\mu,\sigma\rho}\equiv -\kappa (\eta_{\rho\mu}\delta^{0}_{\sigma}-\eta_{\sigma\mu}\delta^{0}_{\rho})\label{Cmixing}.
\ee

\noi In the context of the gauge theory  of Poincar\'e group  the Lorentz curvature is enough to recover Einstein gravity in vacuum, as was pointed out in the subsection 2.2. Therefore, 
in the present model it is enough to consider only the Lorentz and $U(1)$ generalized curvatures in order to construct an electro-gravity theory in the most economical way. The expression of such curvatures reads respectively:

\bea
F^{(\epsilon\rho)}_{\mu\nu}&=&A^{(\epsilon\rho)}_{\mu,\nu}-A^{(\epsilon\rho)}_{\nu,\mu}-
\eta_{\theta\sigma}(A^{(\epsilon\theta)}_{\mu}A^{(\sigma\rho)}_{\nu}-
A^{(\epsilon\theta)}_{\nu}A^{(\sigma\rho)}_{\mu})\,,\nn\\
F^{(\Phi)}_{\mu\nu}&=&A^{(\Phi)}_{\mu,\nu}-A^{(\Phi)}_{\nu,\mu}-
\frac{1}{2}C^{\Phi}_{\epsilon,\theta\rho}(A^{(\epsilon)}_{\mu}A^{(\theta\rho)}_{\nu}-
A^{(\epsilon)}_{\nu}A^{(\theta\rho)}_{\mu})\,\nn\\
&=&A^{(\Phi)}_{\mu,\nu}-A^{(\Phi)}_{\nu,\mu}+
\kappa \eta_{ij}(A^{(j)}_\mu A^{(0i)}_\nu-
A^{(j)}_\nu A^{(0i)}_\mu)\,,
\eea

\noi where $\eta_{ij}$ is the Minkowski metric tensor and 
the latin indices $i,j$ run from 1 to 3 and we recall that
$A^{(\epsilon)}_{\theta}\equiv q^{\nu}_{\theta}\mathcal{A}^{(\epsilon)}_{\nu}=q^{\nu}_{\theta}(k^{\epsilon}_{\nu}-\delta^{\epsilon}_{\nu})=\delta^{\epsilon}_{\theta}-q^{\epsilon}_{\theta}$.


The standard Einstein-Maxwell theory can be described by the gauge theory
associated with the direct product of the Poincar\'e and $U(1)$ groups. But in our present approach corresponding to the central extension the $U(1)$ gauge potential is no longer the usual electromagnetic field $A^{(elec)}_\mu$ in the presence of a gravitational field; rather $A^{(\Phi)}_\mu$  must contain it at zero order in the coupling constant $\kappa$ to account for the limit of the theory without mixing, i.e.

\bea
A^{(\Phi)}_\mu=A^{(elec)}_\mu+\kappa B^{(grav)}_\mu \, .
\eea

\noi In this expression $B^{(grav)}_\mu$ is an ``electromagnetic'' contribution of pure gravitational origin (note that  $B^{(grav)}_\mu$ must be a function of the gravitational potentials). The theory can be developed working up to first order in $\kappa$ and this is, in fact, a good approximation to the problem due to the small value of the coupling constant $\kappa$ (it should be expected that
$|\kappa q|\leq m_{electron}$ and therefore $\kappa$ would result  
to be $\leq 6\times 10^{-12}\; \hbox{Kg/C}$
\cite{egmixing})\footnote{The maximum supposed value for $\kappa$ would 
correspond to the mass-charge relation of the electron. In this case, the 
physical content of the module of  
$\lambda_\mu$ would be essentially the quotient of
coupling constants (gravitational and
electromagnetic ones). 
This is in fact a feature of unified (gauge) theories, for example, in the electro-weak theory the tangent of the Weinberg angle gives precisely
the relation between the isospin and hypercharge coupling constants.}.

Hence, the curvature associated with $U(1)$ can be decomposed into two pieces:
the usual electromagnetic curvature in a gravitational background $F^{(elec)}_{\mu\nu}$ added to a contribution 
constructed from the
gravitational potentials 
$F^{(grav)}_{\mu\nu}$, i. e.

\bea
F^{(\Phi)}_{\mu\nu}=F^{(elec)}_{\mu\nu}+\kappa F^{(grav)}_{\mu\nu}
\eea

\noi with

\bea
F^{(elec)}_{\mu\nu}&=&A^{(elec)}_{\mu,\nu}-A^{(elec)}_{\nu,\mu} \, , \nn\\
F^{(grav)}_{\mu\nu}&=&B^{(grav)}_{\mu,\nu}-B^{(grav)}_{\nu,\mu}+
 \eta_{ij}(A^{(j)}_\mu A^{(0i)}_\nu-
A^{(j)}_\nu A^{(0i)}_\mu)\,.
\eea

\noi As a result we propose that the field $B^{(grav)}_\mu$ could be responsible for some electromagnetic force associated with very massive rotating systems, as
$A^{(0i)}_\mu$ is somehow related to ``Coriolis-like forces''\footnote{Note
that the Lorentz potentials 
$A^{(0i)}_\mu$
can be related to the components 
$\Gamma^i_{00},\Gamma^i_{0k},\Gamma^i_{jk}$ of the
Christoffel symbols 
which produce a Coriolis-like force on a particle in a constant gravitational
field \cite{landauclassical}.}.

\noi The simplest electro-gravitational
gauge invariant Lagrangian density for the free compensating fields in our model has the form:

\bea
L_0&\sim&\Lambda (\mathcal{F}^{(\Phi)}_{\mu\nu}\mathcal{F}^{(\Phi)\mu\nu}+\mathcal{F}^{(\mu\nu)}_{\mu\nu})\,\nn\\
&=&\Lambda (g^{\mu\sigma}g^{\nu\rho}F^{(\Phi)}_{\mu\nu}F^{(\Phi)}_{\sigma\rho}+
k^\sigma_\mu k^\rho_\nu F^{(\mu\nu)}_{\sigma\rho}), \label{mar}
\eea

\noi where 
$\mathcal{F}^{(\Phi)\mu\nu}\equiv \mathcal{F}^{(\Phi)}_{\mu\nu}\eta^{\sigma\mu}\eta^{\rho\nu}$, $g^{\sigma\rho}=k^\sigma_\mu k^\rho_\nu \eta^{\mu\nu}$
and $\Lambda=det(q^\nu_\mu)$.

The Euler-Lagrange motion equations read:

\be
(1):\;\;\;\;\; \frac{\partial L_{0}}{\partial A^{(\nu\rho)}_{\mu}}-\frac{\partial}{\partial x^{\sigma}}\left(\frac{\partial L_{0}}{\partial A^{(\nu\rho)}_{\mu,\sigma}}\right)=0\;\;\Rightarrow \label{(1)} 
\label{(1)}
\ee

\[C^{\Phi}_{\sigma,\epsilon\theta}A^{(\sigma)}_{\nu}F^{(\Phi)\mu\nu}+k^{\mu}_{\rho}T^{\rho}_{\epsilon\theta}-k^{\mu}_{\theta}T^{\rho}_{\epsilon\rho}+k^{\mu}_{\epsilon}T^{\rho}_{\theta\rho}+
(k^{\mu}_{\rho}k^{\nu}_{\theta}-k^{\mu}_{\theta}k^{\nu}_{\rho})
A^{(\rho}_{\;\;\;\epsilon)\nu}-(k^{\mu}_{\epsilon}k^{\nu}_{\rho}+k^{\mu}_{\rho}k^{\nu}_{\epsilon})A^{(\rho}_{\;\;\;\theta)\nu}=0\,,\]

\noi where
$F^{(\Phi)\mu\nu}=F^{(\Phi)}_{\rho\lambda}g^{\rho\mu}g^{\lambda\nu}\,,$
$g^{\rho\mu}=k^{\rho}_{\sigma}k^{\mu}_{\theta}\eta^{\sigma\theta}\,,$
$A^{(\mu}_{\;\;\;\nu)\sigma}\equiv \eta_{\nu\rho}A^{(\mu\rho)}_{\sigma}$ and 
$T^{\rho}_{\epsilon\theta}=q^{\rho}_{\mu}(k^{\mu}_{\epsilon,\tau}k^{\tau}_{\theta}-k^{\mu}_{\theta,\tau}k^{\tau}_{\epsilon})\;;$
\be
(2):\;\;\;\;\;  \frac{\partial L_{0}}{\partial A^{(\Phi)}_{\mu}}-\frac{\partial}{\partial x^{\sigma}}\left(\frac{\partial L_{0}}{\partial A^{(\Phi)}_{\mu,\sigma}}\right)=0\;\;\Rightarrow \label{(2)}
\ee

\[\frac{d}{dx^{\sigma}}(\Lambda F^{(\Phi)\mu\sigma})=0\;;\]

\be
(3):\;\;\;\;\;\;\;  \frac{\partial L_{0}}{\partial k^{\mu}_{\nu}}-\frac{\partial}{\partial x^{\sigma}}\left(\frac{\partial L_{0}}{\partial k^{\mu}_{\nu,\sigma}}\right)=0\;\;\Rightarrow \label{(3)}
\ee

\[F^{(\nu\sigma)}_{\mu\sigma}-\frac{1}{2}\delta^{\nu}_{\mu}F^{(\sigma\lambda)}_{\sigma\lambda}=\mathcal{T}^{\nu}_{\mu}\,,\]

\noi where

\[\mathcal{T}^{\nu}_{\mu}\equiv \mathcal{T}^{\nu(mix)}_{\mu}+
\mathcal{T}^{\nu(\Phi)}_{\mu}\,.\]

\noi The tensor
$\mathcal{T}^{\nu(\Phi)}_{\mu}\equiv -F^{(\Phi)\nu}_{\sigma}F^{(\Phi)\sigma}_{\mu}+\frac{1}{2}\delta^{\nu}_{\mu}F^{(\Phi)}_{\sigma\lambda}F^{(\Phi)\sigma\lambda}\,,$ generalizes the energy-momentum tensor corresponding to the electromagnetic field in a gravitational field (with $F^{(\Phi)\nu}_{\sigma}=g^{\lambda\nu}F^{(\Phi)}_{\sigma\lambda}$) and the piece 
$\mathcal{T}^{\nu(mix)}_{\mu}=\frac{1}{2}C^{\Phi}_{\mu,\theta\epsilon}q^{\nu}_{\rho}F^{(\Phi)\rho\tau}A^{(\theta\epsilon)}_{\tau}\;$
is completely new and arises as a direct consequence of the mixing of the space-time and internal symmetries.

In order to proceed  further in the understanding of the proposed model we shall consider the effects of the mixing of gravity and electromagnetism in the 
``geodesic'' motion. Let us consider a spinless particle of mass m,
 momentum
$p_{\mu}(=mu_{\mu}=m\frac{dx_{\mu}}{d\tau})$ and charge $e$.
According to the (Generalized) Minimal Coupling Principle,
the Lagrangian of the free particle 
\bea
\mathcal{L}_{particle}=\frac{1}{2m}p_{\mu}p_{\nu}\eta^{\mu\nu}
\eea
 
\noi must be replaced by the modified Lagrangian where $p_\mu\rightarrow k^\nu_\mu (p_\nu -eA^{(\Phi)}_\nu)=
k^\nu_\mu (p_\nu-eA^{(elec)}_\nu-\kappa e B^{(grav)}_\nu)
$:

\bea
{\widehat{\mathcal{L}}}_{particle}=
\frac{1}{2}m u^{\mu}u^{\nu}g_{\mu\nu}-eu^{\mu}A^{(elec)\nu}g_{\mu\nu}
-\kappa e u^{\mu}B^{(grav)\nu}g_{\mu\nu}
\,,
\eea where we have already neglected
the misleading term 
$\frac{e^{2}}{2m}A^{(\Phi)\mu}A^{(\Phi)\nu}g_{\mu\nu}$
\footnote{We recall that already in the standard formulation of the Lorentz force
in a gravitational field (without mixing) the interaction Lagrangian 
$L_{int}$, among some other requisites, must be linear in the charge of the particle and in the electromagnetic 
potential
to account for the Lorentz invariance of $\gamma L_{int}$ (with
$\gamma\equiv (1-(\frac{u}{c})^2)^{-\frac{1}{2}}$) as a consequence
of the requirement of Lorentz invariance of the action integral written in 
terms of the proper time $\tau$ \cite{jackson}.}
, which, by the way, does not appear when working directly with 
the Poincar\'e-Cartan form instead of the Lagrangian. We also consider (\ref{mar}) as the 
Lagrangian density for the free compensating fields.
As regards the interaction between a particle and a field, in general, it is required to 
distinguish between the coordinates $y^{\sigma}$ where the fields are evaluated and the 
coordinates $x^{\sigma}$ for the particle.
In ${\widehat{\mathcal{L}}}_{particle}$ the fields are evaluated at the position of the particle, 
where the interaction occurs, but in $\mathcal{L}_{0}$ the fields are evaluated at $y^{\sigma}$.

The equation for the particle,
$\frac{\partial {\widehat{\mathcal{L}}}_{particle}}{\partial x^{\sigma}}-\frac{d}{d\tau}\left(\frac{\partial {\widehat{\mathcal{L}}}_{particle}}{\partial u^{\sigma}}\right)=0$
 results in the usual motion equation for a particle in the presence of 
both gravitational and electromagnetic fields, with an additional Lorentz-like
force (proportional to $\kappa e$) generated by 
the gravitational potentials, i.e.

\be
g_{\mu\sigma}\frac{du^{\mu}}{d\tau}=-u^{\mu}u^{\nu}\Gamma^{(L-C)}_{\mu\nu,\sigma}-\frac{e}{m}u^{\mu}F^{(elec)}_{\mu\sigma}-\frac{\kappa e}{m}u^{\mu}(\partial_{\sigma}B^{(grav)}_{\mu}-\partial_{\mu}B^{(grav)}_{\sigma})\,.\label{motion}
\ee

\noi with $\Gamma^{(L-C)}_{\mu\nu,\sigma}=\frac{1}{2}
(\partial_\mu g_{\nu\sigma}+\partial_\nu g_{\mu\sigma}-
\partial_\sigma g_{\mu\nu})$ being the Levi-Civita connection associated with the metric $g_{\mu\nu}=q_\mu^\rho q_\sigma^\nu \eta_{\rho\nu}$.

\noi Considering the non-relativistic limit ($c\rightarrow \infty$) 
on the Poincar\'e group (stated as an Inonu-Wigner Lie algebra contraction \cite{mixcorto}) the explicit form of the field $B^{(grav)}_\mu$
 is then very simple ($h^i\equiv g^{0i}-\eta^{0i}$): 
\bea
B^0&=&-\frac{\vec{h}^2}{8}\nn\\
 B^i&=&-\frac{h^i}{2}\,.
\eea

It must be remarked, however, that in any case we do not refer to a new force but, just, a mixing of 
interactions, so that the number of field degrees of freedom are the same that in the 
$\kappa\rightarrow 0$ limit. 

Some final comments are in order: firstly, since the present theory has been formulated on symmetry grounds, 
it could be possible 
to attempt the
quantization on the basis of the Group Approach to Quantization. 
With regard this question the purpose of the GAQ treatment for the quantization of gravity would consist in restricting ourselves
to a subgroup of the supposed symmetry group of gravity. Thus using this subgroup to parametrize the corresponding solution submanifold 
(Schwarzschild-like solution, for instance) one could manage to describe the theory with a lower 
number of parameters (even finite) in a non-perturbative framework, then avoiding 
renormalizability problems.  
Secondly, the unification of gravity and electromagnetism here proposed can be inmediately generalized to the rest of interactions once the group $U(1)$ is considered as a subgroup
 of $(SU(2)\otimes U(1))/Z_{2}$, $SU(5)$ or any other 
``grand unification group''. Finally, we also remark that the semidirect
product of the diffeomorphism group of the space-time and the gauge group,
$Diff(M)\otimes_S G(M)$, 
provides an extra natural mixing between gravity and the rest of (internal) interactions, although 
maybe less drastic in phenomenological terms than the mixing proposed here. In fact, in the 
case of electromagnetism, the semi-direct 
action of the group of diffeomorphisms on the gauge group $U(1)(M)$ would account 
for diagrams in which photons and gravitons produce gravitons. Thus this mixing would result in a new modified dispersion relation between 
gravitons and photons. However in the context of gauging the central extension of Poincar\'e group by $U(1)$, diagrams in which two gravitons provide one photon 
would enter the theory. In such a case the production of photons in the absence of electrically 
charged sources would be expected.

\section*{Acknowledgments}

The authors wish to thank Carlos Barcel\'o for useful discussions and for reading the manuscript. E. S.-S. is grateful to B. N. Frolov for very valuable discussions and suggestions and thanks the Department of Physics of Moscow State Pedagogical University for its hospitality.

\end{document}